\documentclass[12pt]{article}
\usepackage{latexsym}
\usepackage{amsmath,amsbsy,amssymb}

\usepackage{curves}
\usepackage{epic}
\usepackage{eepic}
\usepackage{epsfig}

\newcommand{\dd}{\mbox{\rm d}}

\newcommand{\tl}{\tilde}

\newcommand{\be}{\begin{equation}}
\newcommand{\bear}{\begin{eqnarray}}

\newcommand{\ear}{\end{eqnarray}}
\newcommand{\ee}{\end{equation}}
\newcommand{\lbl}{\label}
\newcommand{\bi}{\bibitem}
\newcommand{\ci}{\cite}

\newcommand{\vs}{\vspace}
\newcommand{\hs}{\hspace}

\begin{document}

\begin{center}

\

\vs{.8cm}

\baselineskip .7cm

{\bf \Large Extra Time Like Dimensions, Superluminal Motion, and
Dark Matter } \\

\vs{2mm}

\baselineskip .5cm
Matej Pav\v si\v c

Jo\v zef Stefan Institute, Jamova 39,
1000 Ljubljana, Slovenia

e-mail: matej.pavsic@ijs.si

\vs{3mm}

{\bf Abstract}
\end{center}

\baselineskip .43cm
{\small

We show that the superluminal speeds of the muon neutrinos observed
in the OPERA experiment can be explained within a relativity theory
with extra time like dimensions. In addition, such theory predicts,
the existence of dark matter.}

\hs{7mm}

\baselineskip .55cm

\section{Introduction}

The recent OPERA finding of faster-than-light muon neutrinos\,\ci{Opera}, if
confirmed, and if it cannot be explained within the usual physics, will
require a reformulation or a generalization of the theory of relativity.
There already exist several theories that allow for superluminal motion,
and they all contain the Einstein's special relativity as a limiting case,
for instance:

\ I. The extended special relativity with superluminal
transformations\,\ci{PavsicExtend,RecamiRiv1,RecamiRiv2}. The latter theory joins
the Bilaniuk-Despande-Sudarshan proposal of tachyons\,\ci{Sudarshan} with
the principle of relativity.

II. Higher dimensional spaces, $M_{p,q}$, with extra time like and space like
dimensions (signature $(p,q)$)\,\ci{ExtraTime,PavsicCausality}.
In a subspace with signature $(1,3)$
(the Minkowski spacetime $M_{1,3}$), particles can move faster than light,
if their worldlines are suitably inclined into the extra time like
dimensions\,\ci{Barashenkov}.  A candidate for such higher dimensional space
is the Clifford space $C_{8,8}$, a manifold whose tangent space at every
point is a Clifford algebra $Cl(1,3)$, generated by basis vectors of $M_{1,3}$.
In the past years an extended relativity theory in Clifford spaces was
developed\,\ci{CastroPavsicReview}--\ci{CliffordSpace}.
As discussed in \ci{PavsicArena,CastroPavsicReview},
one can have tachyonic (faster-than-light) behavior in ordinary
spacetime, $M_{1,3}$, while having {\it non}-tachyonic behavior in $C_{8,8}$.

Besides that, the Stueckelberg theory with an invariant evolution
parameter\,\ci{FanchiRev,PavsicBook}, and the relativity in phase
space\,\ci{PhaseSpace} also predict superluminal motions.

In this Letter I will show that the extended special relativity (Case I)
cannot explain the OPERA result $v/c-1=2.48 \times 10^{-5}$, because this
would require too high neutrino mass. On the other hand, in the presence
of extra time like dimensions (Case II), a particle's speed, as observed in the
ordinary spacetime, can be greater than the
speed of light, regardless of how small the particle's mass is.
Usually, such a possibility of superluminal motion  has been considered
as an argument against extra time like dimensions. But this argument will
no longer hold, if the OPERA result is confirmed.

Moreover, in a space $M_{p,q}$ there exist particle's worldlines that are
invisible to certain classes of observers\,\ci{ColeDark,Barashenkov}.
And vice versa, for a given observer
${\cal O}$ in $M_{p,q}$, associated with a worldline $A$, there exist
a class of worldlines $\{B\}$ that are invisible to ${\cal O}$, because
the light cones originating from the points on $B$ do not intersect 
the observer's worldline $A$. This could be an explanation for dark matter.
But such matter would not be really dark, but only invisible to us.

\section{Faster-than-light motion}
\subsection{Tachyons in the extended special relativity}

The possibility of tachyons was proposed by Bilaniuk et al.\,\ci{Sudarshan}. In 70's the
idea attracted considerable attention and the principle of relativity was extended
to superluminal transformations\,\ci{PavsicExtend,RecamiRiv1,RecamiRiv2}.

A tachyons's energy is given by
\be
    E= \frac{m c^2}{\sqrt{\frac{v^2}{c^2}-1}}
\lbl{2.1}
\ee
From this we find
\be
    \frac{v}{c} = \sqrt{1+\frac{m^2 c^4}{E^2}} \approx 1 + \frac{1}{2}\frac{m^2 c^4}{E^2}
\lbl{2.1a}
\ee
Taking the OPERA result $v/c-1 = 2.48 \times 10^{-5}$ and $E=17$\,GeV, we have
\be
    m c^2 \approx E \sqrt{2(\frac{v}{c}-1)}= 17 \times 10^9 {\rm eV} 
    \sqrt{2 \times 2.48 \times 10^{-5}} = 120\, {\rm MeV} .
\lbl{2.2}
\ee
Since the muon neutrino mass is much lower, it means that the faster-than-light neutrinos
found in the OPERA experiment, cannot be the tachyons of the kind considered
in\,\ci{PavsicExtend}--\ci{Sudarshan}.

On the other hand, the SN1987a neutrinos\,\ci{Supernova} had energy around 20\,MeV and
the velocity $v/c-1 \approx 2 \times 10^{-9}$. From Eq.\,(\ref{2.2}) we then obtain
$m c^2 \approx 9 \times 10^2 {\rm eV}$. This also is too high value for the neutrino mass.

\subsection{Beyond the speed of light in spaces with extra time like dimensions}

The idea that spacetime has equal number of space like and time like dimensions has
been much explored in the past decades\,\ci{ExtraTime,PavsicCausality}.
Let $M_{n,n}$ be a manifold whose
points are described by coordinates $x^a = (t^i,x^i),~i=1,2,...,n$, and the quadratic
form is given by
\be
    \dd s^2 = \eta_{ab} \dd x^a \dd x^b= \dd t^i \dd t_i + \dd x^i \dd x_i
\lbl{2.3}
\ee
with the metric $\eta_{ab} = {\rm diag} (1,1,...,-1,-1,...)$.
We use the units in which the
speed of light, defined according to $c^2 =
(-\dd x^i \dd x_i)/(\dd t^i \dd t_i)$, is $c=1$.

In such spacetime, the transformations that change the sign of $\dd s^2$
(the so called `superluminal transformations' that transform a bradyon
into a tachyon), are real. However, a slower than light particle, a bradyon,
cannot be accelerated beyond the speed of light and become a tachyon, because
of the infinite energy barrier on the light cone in $M_{n,n}$. And yet,
if our physical space is of the type $M_{n,n}$, then faster than light travel
is possible in principle, because in the 4D subspace $M_{1,3}$, an accelerating
particle can become faster than light, without crossing the infinite
energy barrier. A particle does not need to cross the light cone in $M_{n,n}$
in order to become superluminal in $M_{1,3}$. Such a particle is not a
tachyon in $M_{n,n}$, because its worldline is still subluminal with
respect to the light cone in $M_{n,n}$, and the $\dd s^2$ along the worldline
did not change sign.

The action for a massive particle moving in $M_{n,n}$ is
\be
  I[x^a] = M \int \dd \tau \, ({\dot x}^a {\dot x}^b \eta_{ab})^{1/2} ,
\lbl{2.4}
\ee
where ${\dot x}^a \equiv {\dot x}^a (\tau)$ are functions of an arbitrary parameter
$\tau$.

The momentum is
\be
    p^a = \frac{M {\dot x}^a}{({\dot x}^c {\dot x}^d \eta_{cd})^{1/2}}=
    ({\tl p}^i,p^i)
\lbl{2.5}
\ee
and satisfies the constraint
\be
    p^a p^b \eta_{ab} = M^2 .
\lbl{2.6}
\ee
From now on, we take $n=3$.

Since $\tau$ is arbitrary, we may take $\tau = t^1$. Then we have
\be
   p^a = \frac{M {\dot x}^a}{\left (1+({\dot t}^2)^2+({\dot t}^3)^2
 - ({\dot x})^1- ({\dot x})^2-({\dot x})^3 \right )^{1/2}} ,
\lbl{2.7}
\ee
where
\be
     v^2 \equiv ({\dot x}^1)^2+ ({\dot x}^2)^2 +({\dot x}^3)^2 < 
    1 +({\dot t}^2)^2+({\dot t}^3)^2
\lbl{2.8}
\ee
and ${\dot x}^a = \dd x^a/\dd t^1$. From the latter equation it follows
that the spatial speed, $v$, can exceed the speed of light, $c=1$, if
$w^2 \equiv ({\dot t}^2)^2+({\dot t}^3)^2 > 0$, while still satisfying
the condition $1-v^1+w^2 >0$.

We define the first component as energy\footnote{We assume that $\tau=t^1$
is the proper time of the observer ${\cal O}$ who measures the
momentum $p^a=(\tl p^1, \tl p^2, \tl p^3,p^1,p^2,p^3)$ of the particle.
A particle that is at rest with respect to ${\cal O}$ has then only the
first component of momentum different from zero: $p'^a=(\tl p^1,0,0,0,0,0)$.
Therefore, for the observer ${\cal O}$, the $\tl p'^1$ and $\tl p^1$ are
the energies of the particles. }
\be
    \tl p^1 \equiv E = \frac{M}{(1-v^2+w^2)^{1/2}} ,
\lbl{2.9}
\ee
from which it follows
\be
   w = \sqrt{v^2-1+\frac{M^2}{E^2}} .
\lbl{2.10}
\ee
The muon neutrinos of the OPERA experiment had the average energy
$E=17$\,GeV and the velocity $v=1+2,48 \times 10^{-5}$. Using those data
in Eq.\,(\ref{2.10}), we obtain
\be
     w\approx \sqrt{2 (v-1)} = 7,04 \times 10^{-3} ,
\lbl{2.11}
\ee
to which there corresponds the velocity $2.11 \times 10^6\,m/s$. This
is the velocity in the direction of the time like dimensions
$t^2$ and $t^3$ that gives the superluminal velocity $v$ of the muon neutrino.

From Eq.\,(\ref{2.9}) we also have
\be
   E^2 = \frac{M^2}{1-v^2+w^2} = \frac{M^2}{1-v^2}\frac{1-v^2}{1-v^2+w^2}
     = \frac{m^2}{1-v^2} ,
\lbl{2.12}
\ee
where
\be
   m^2 = M^2 \frac{1-v^2}{1-v^2+w^2}
\lbl{2.13}
\ee
is the effective mass in 4D Minkowski space spanned over $(t^1,x^1,x^2,x^3)$.
The relation (\ref{2.13}) is well-known from Kaluza-Klein theories. It is
a consequence of the relations
\be
    \dot x^a \dot x^b \eta_{ab} = \dot x^\mu \dot x^\nu \eta_{\mu \nu} +
    {\dot x}^{\bar a} {\dot x}^{\bar b} g_{{\bar a}{\bar b}} 
\lbl{2.14}
\ee
and
\be
    M^2 = p^a p^b \eta_{ab} = p^\mu p^\nu \eta_{\mu \nu}+ p^{\bar a} p^{\bar b} 
    g_{{\bar a}{\bar b}} ,
\lbl{2.15}
\ee
where $\dot x^\mu = (\dot t^1,\dot x^1,\dot x^2,\dot x^3)\equiv
(\dot x^0,\dot x^i)$ is the 4-velocity, $p^\mu=
(\tl p^1,p^1,p^2,p^3) \equiv (p^0,p^i)$ the 4-momentum, whereas
${\dot x}^{\bar a} = (\dot t^2,\dot t^3)$ and $p^{\bar a} =(\tl p^2,\tl p^3)$
are extra components of velocity and momentum, respectively.
Multiplying Eq.\,(\ref{2.14}) with $M^2$, using $M^2=p^a p^b \eta_{ab}$,
$m^2=p^\mu p^\nu \eta_{\mu \nu}$, and Eq.\,(\ref{2.5}), and identifying
$\dot x^i \dot x_i = -v^2$, ${\dot x}^{\bar a} {\dot x}_{\bar a}
=(\dot t^2)^2 + (\dot t^3)^2 \equiv w^2$, we obtain Eq.\,(\ref{2.13}).
Let us also identify $p^{\bar a} p^{\bar b} g_{\bar a \bar b} =
({\tl p}^2)^2 +({\tl p}^3)^2 \equiv \tl q^2$. Then Eq.\,(\ref{2.15})
can be written as
\be
   M^2 = m^2 + \tl q^2
\lbl{2.16}
\ee

If $w^2 \neq 0$, then it can be $v^2 > 1+w^2 > 1$. The 4D mass squared (\ref{2.13}) is
then negative, $m^2 <0$. From Eq.\,(\ref{2.12}) we obtain
\be
   v^2-1 = \frac{-m^2}{E^2}= \frac{\tl q^2 -M^2}{E^2} .
\lbl{2.17}
\ee
So we have
\be
   \sqrt{-m^2} = \sqrt{\tl q^2 - M^2} \approx E \sqrt{2 (v-1)} =
   17\,{\rm GeV} \sqrt{2.48 \times 10^{-5}} = 120\,{\rm MeV} .
\lbl{2.19}
\ee
The same
equation has been derived in Ref.\,\ci{PhaseSpace} by a different procedure,
starting from the dispersion relations and the expression for the group
velocity. It was also observed that the above mass is close to the muon mass
$m_\mu = 105.7$\,MeV.  Whether or not this is a coincidence has to be
found out by further investigations.

We see that in this model the quantity $m^2= M^2-{\tl q}^2$ is the
effective 4D mass squared of the muon neutrino. We envisage that at high energies,
in the collision processes producing mesons $\pi^+$, $K^+$ that
subsequently decayed into $\mu^+$ and $\nu_\mu$, the particles acquired
not only the ordinary spatial momenta $p^1,p^2,p^3$, but also the
extra momenta $\tl p^2, \tl p^3$, which, according to Eq.\,(\ref{2.19}),
contributed to the effective 4D mass of $\nu_\mu$. According to this
theory, the 4D masses of $\nu_\mu$, $\nu_\tau$ and $\nu_e$ depend on the
conditions of the process in which they are produced. In a decay process
of a low energy pion, $\pi^+ \rightarrow \pi^0 + \mu^+ + \nu_\mu$,
with the pion momentum $p^a=(\tl p^1,0,0;p^1,p^2,p^3)$, the outgoing
$\mu^+$ and $\nu_\mu$ have low spatial momenta, and also low extra
momenta $\tl p^2, \tl p^3$, and they have thus low 4D masses, $m \approx M$.

Neutrino oscillations give us the information about the differences\footnote{
Here we distinguish the weak eigenstates, $e$, $\mu$, $\tau$, from
the mass eigenstates, $e'$, $\mu'$, $\tau'$, by using prime, though more
common notation is $1, 2, 3$.}
$m_{\nu_{\mu'}}^2-m_{\nu_{\tau'}}^2$, $m_{\nu_{\mu'}}^2-m_{\nu_{e'}}^2$, etc.,
but not about the  values $m_{\nu_{\mu'}}^2 = 
M_{\nu_{\mu'}}^2- {\tl q^2}_{\nu_{\mu'}}$, $~m_{\nu_{\tau'}}^2 = 
M_{\nu_{\tau'}}^2-{\tl q^2}_{\nu_{\tau'}}$, and $m_{\nu_{e'}}^2 = 
M_{\nu_{e'}}^2-{\tl q^2}_{\nu_{e'}}$. In the above differences, the extra
momenta cancel out, if they are all equal to each other. This is indeed the
case, because the 6-momentum must be conserved when one kind of neutrino
oscillates into another one in the absence of any external interactions.

\section{Invisible particles}

In a world with multi-time like dimensions, a luminous body is not visible
for all observers\,\ci{ColeDark,Barashenkov}. For instance, if an
observer worldline $A$ is displaced ``sidewise" in multi-time with
respect to a body's wordlline $B$ (see Fig.\,1), then $B$ can be
invisible to $A$, and vice versa. This is so, because the light cones
originating\footnote{In Refs.\,\ci{ColeDark,Barashenkov}, the authors
consider the light cones that originate from the observer. 
Therefore, strictly speaking, they find under which conditions the
observer is not seen by the particle. But actually, we are interested in the
opposite, namely, when the particle (matter) is not seen by the observer.
Therefore, I consider the light cones originating form the particle worldline.
The procedure that I then employ is just a straightforward application of
analytical geometry.}
 from $B$ do not intersect the worldline $A$. A light cone
is given by
\be
    (t^i-t_B^i)^2 - (x^i-x_B^i)^2 = 0.
\lbl{3.1}
\ee
Formally, the worldlines of the observer $A$ and the luminous source $B$
are
\be
   t^i = t_A^i (\tau)~,~~~x^i = x_A^i (\tau)~;~~~
   t^i = t_B^i (\tau_0)~,~~~x^i = x_B^i (\tau_0) .
\lbl{3.2}
\ee
In particular, let the worldline of $A$ be
\be
    A:~~~t^1 =\tau~,~~~t^2=t_A^2 \neq 0~,~~~t^3=0~,~~~x^1=x_A^1 \neq 0~,
    ~~~x^2=0~,~~~x^3=0
\lbl{3.3a}
\ee
\be
    B:~~~t^1 =\tau_0~,~~~t^2=0~,~~~t^3=0~,~~~x^1= 0~,~~~x^2=0~,~~~x^3=0 .
\lbl{3.3b}
\ee
We thus have $t_B^1 (\tau_0) = \tau_0$, which is the time of emission
of light.

\setlength{\unitlength}{.8mm}
\begin{picture}(150,90)(-75,-20)

\put(0,0){\vector(1,0){62}}
\put(0,0){\vector(0,1){62}}
\put(0,0){\vector(-2,-1){45}}

\newsavebox{\cone}
\savebox{\cone}(0,0,)[l]{

\put(0,0){\line(1,1){17}}
\put(0,0){\line(1,-1){17}}
\renewcommand{\xscale}{0.5}
\renewcommand{\xscaley}{0.5}
\renewcommand{\yscale}{1}
\renewcommand{\yscalex}{-1}
\put(19.8,0){\bigcircle{25}}
}

\renewcommand{\xscale}{1}
\renewcommand{\xscaley}{-1}
\renewcommand{\yscale}{0.55}
\renewcommand{\yscalex}{0.55}
\put(4,-12){\bigcircle{2}}

\put(0,0){\usebox{\cone}}
\put(0,38){\usebox{\cone}}

\multiput(4,-12)(3.15,2.1){6}{\line(3,2){1.5}}
\multiput(4,-12)(-3.2,0){9}{\line(-3,0){1.5}}
\put(-25,-17){$t_A^2$}
\put(20,-4.7){$x_A^1$}

\put(64,1){$x^1$}
\put(1,64){$t^1$}
\put(-49,-21){$t^2$}
\put(-5.3,20){$B$}
\put(5,16){$A$}
\put(-5.3,37){$\tau_0$}

\thicklines
\put(0.1,0){\line(0,1){55}}
\multiput(0,0)(0,-4){5}{\line(0,-1){2}}
\put(4,-12){\line(0,1){62}}
\multiput(4,-12)(0,-4){4}{\line(0,-1){2}}

\end{picture}

\begin{figure}[h]
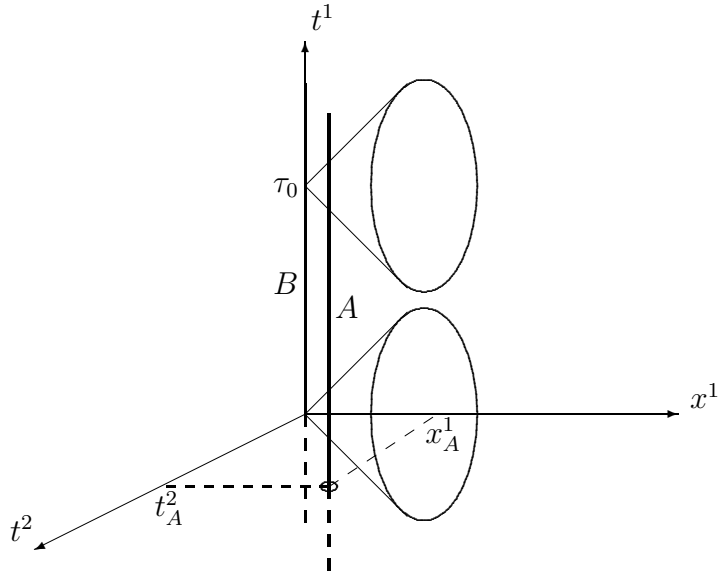

\caption{ \small The light cones originating from the worldline $B$, traced by
a luminous source, never intersect the observer's worldline $A$ that is
parallelly displaced with respect to $B$ along the direction of $t^2$. Therefore,
this source is invisible to the observer.}

\end{figure}

A light cone, emerging from a point on the worldline $B$ is
\be
    (t^1-\tau_0)^2+(t^2)^2+(t^3)^2 - (x^1)^2 - (x^2)^2 - (x^3)^2 = 0 .
\lbl{3.4}
\ee
Inserting the values for $t^i$, $x^i$ of Eq.\,(\ref{3.3a}) into
Eq.\,(\ref{3.4}), we obtain
\be
    (\tau -\tau_0)^2+(t_A^2)^2 - (x_A^1)^2 = 0 .
\lbl{3.5}
\ee
If $(t_A^2)^2 - (x_A^1)^2 > 0$, then the system of equations (\ref{3.3a})
and (\ref{3.4}) has no solution, since by our assumption all coordinates
are real. Therefore, the light cone (\ref{3.4}) and the worldline $A$ have 
no common point, i.e., they do not intersect. This is true for any value of
the parameter $\tau_0$, and hence for any point on the source worldline $B$.

In general, the observer's worldline $A$ need not be parallel to that of
the source. Let $A$ be given by
\be
   t^1 = u^1 \tau~,~~~t^2=u^2 \tau+t_A^2~,~~~t^3=0~,~~~x^1=x_A^1 \neq 0~,
    ~~~x^2=0~,~~~x^3=0,
\lbl{3.6}
\ee
where $(u^1)^2+(u^2)^2=1$.
For the source world line we keep Eq.\,(\ref{3.3b}). Inserting (\ref{3.6})
into the light cone equation (\ref{3.4}), we have
\be
    (u^1 \tau -\tau_0)^2+ (u^2 \tau+t_A^2)^2 - (x_A^1)^2 = 0 .
\lbl{3.7}
\ee
From the latter equation we obtain
\be
    \tau= -(u^2-u^1 \tau_0) \pm \sqrt{(u^2 t_A^2-u^1 \tau_0)^2-\left ( \tau_0^2
    +(t_A^2)^2 - (x_A^1)^2 \right )}
\lbl{3.7a}
\ee    
This has a solution, if the discriminant is positive:
\be
(u^2 t_A^2-u^1 \tau_0)^2-\left ( \tau_0^2 +(t_A^2)^2 - (x_A^1)^2 \right ) >0 .
\lbl{3.7b}
\ee
This holds for a certain interval of the $\tau_0$ values that can be
found by solving the quadratic equation
\be
    -(u^2)^2 \tau_0^2 - 2 u^1 u^2 t_A^2 \tau_0 +
    (t_A^2)^2 \left ( (u^2)^2-1 \right )+(x_A^1)^2 = 0,
\lbl{3.8}
\ee
which gives
\be
    \tau_0 = \frac{-u^1 t_A^2 \mp x_A^1}{u^2} .
\lbl{3.9}
\ee
Assuming $x_A^1 >0$, the source $B$ is visible for the observer $A$
within the interval
\be
    \frac{-u^1 t_A^2 -x_A^1}{u^2} < \tau_0 < 
    \frac{-u^1 t_A^2 +x_A^1}{u^2}
\lbl{3.10}
\ee
Recall that $\tau_0 = t_B^1$ is the time of emission.
The corresponding time of detection $\tau =\sqrt{(t^1)^2+(t^2)^2}$
can be calculated from Eq.\,(\ref{3.7a}).
Since detection of light by the observer $A$ must be after its emission
by the source $B$, we require $\tau > \tau_0$. The latter requirement
determines the sign in Eq.\,(\ref{3.7a}). Thus, if $u^1 >0$, one must
take the positive sign in fron of the square root.
Outside the interval (\ref{3.10}), the source is invisible for $A$.

Dark matter
could be an effect of a higher dimensional spacetime with extra time like
dimensions. Certain matter, whose worldlines are parallelly displaced
in multi-time with respect to our worldline, is dark all the time.
Those worldlines that are inclined in multi-time, are visible to us for
a certain time period, and invisible before and after that period.
For astrophysical objects such period is very long, because the distance
from the source, $x_A^1$, occurring in Eq.\,(\ref{3.10}), is very large.
Therefore, it is unlikely
that we will suddenly see the appearance of a new object or the disappearance
of an existing one. Moreover, since the object consists of many particles
whose world lines have different directions in multi-time\footnote{This
is a very natural assumption, because the object's worldlines also have
different directions in 3-space, i.e, different velocities (not all particle
forming the object move in the same direction and with the same speed).},
the appearance or disappearance of such object would not be sudden,
but gradual. From Eq.\,(\ref{3.9}) we estimate the transition
period to be $\Delta \tau_0 \approx (x_A^1/u^2)(\Delta u^2/u^2)$, where
$\Delta u^2$ is the average spread of $u^2$. Taking, e.g.,
$x_A^1 \approx10^5$ light years, $u^2 \approx 10^{-3}$, and
$\Delta u^2/u^2 \approx 10^{-5}$, we obtain $\Delta \tau_0 \approx 10^3$
years. Thus, on the Earth, we would just see a faint astrophysical
object, and centuries or millennia later the future astronomers will eventually see
that those objects have become slightly fainter or slightly more luminous.

\section{Conclusion}

Spaces $M_{n,n}$ with multiple time dimensions admit apparent superluminal
motion in the 4-dimensional spacetime $M_{1,4}$. We have investigated
whether this could be an explanation for the superluminal neutrinos
found in the OPERA experiment. Using the generalized expression for
a particle's energy, we have found\footnote{See also an alternative
procedure of Ref.\,\ci{PhaseSpace}.} that a 17 GeV muon neutrino traveling
with the speed $v/c=1+2.48 \times 10^{-5}$ has an imaginary effective 4D mass
of 120 MeV. The latter mass is an invariant under Lorentz transformations
in $M_{1,4}$, but not in $M_{n,n}$, and is the sum of two contributions.
One contributions comes from the invariant mass $M$ in $M_{n,n}$, whereas the
other contribution comes from the momenta in the extra time like dimensions.
The mass $M$ is the true mass of neutrino, and it can be small.
Our conclusion is that the OPERA superluminal muon neutrinos can be
explained in terms of the relativity in $M_{n,n}$. In this paper
we have considered a generic space with time like dimensions. In particular,
a candidate for such space is the Clifford space\,\ci{CliffordSpace}.

We also point out, and rederive in a different way, a known
result\,\ci{ColeDark,Barashenkov} that the special relativity in $M_{n,n}$
predicts dark matter. Such an extended relativity is thus not only
able to explain superluminal neutrinos, but also dark matter. If the theory
is generalized to curved spaces, then, \` a la Kaluza-Klein, it can
describe other interactions, besides the 4D gravity. This possibility
has been explored within the context of Clifford
space\,\ci{PavsicKaluza,PavsicE8}.

A great stumbling block against the acceptance of the possibility of
superluminal velocities is the issue of causality. But if the superluminal
velocities found in the OPERA experiment are confirmed\footnote{If the OPERA
result turns out to be explicable in more conventional terms, there still
remains a possibility that, in the future, superluminal motions will be observed
in a different experimental setup.}
by other independent
experiments, this will pave a way for a reconsideration
of the usual causality arguments. In the literature there already exist
alternative explanations\,\ci{PavsicCausality,PavsicBook}.
Causality paradoxes of tachyons can be resolved\,\ci{PavsicCausality,PavsicBook}
in the same way as David Deutsch resolved\,\ci{Deutsch} the time travel
paradoxes of wormholes: By considering multiverse and the Everett
interpretation of quantum mechanics\,\ci{Everett}. 
The existence of superluminal particles, as well as the existence
of time travel, is in agreement with the Everett
interpretation, but not with the other known
interpretations of quantum mechanics.

\vs{4mm}

\centerline{\bf Acknowledgment}

This work has been supported by the Slovenian Research Agency.

\end{document}